\documentclass[twocolumn,floatfix, nofootinbib,prd,preprintnumbers,showpacs,showkeys,superscriptaddress]{revtex4-1}

\usepackage[mathscr]{euscript}
\usepackage{amsmath}
\usepackage{graphicx}
\usepackage{dcolumn}
\usepackage{bm}
\usepackage{epsfig}
\usepackage{amssymb,latexsym,mathrsfs}
\usepackage{graphicx}
\usepackage{color}
\usepackage{hyperref}
\usepackage{float}
\usepackage{diagbox}

\usepackage{tikz}
\usepackage[compat=1.1.0]{tikz-feynman}

\usepackage{amsmath}
\newcommand*\diff{\mathop{}\!\mathrm{d}}
\newcommand*\Diff[1]{\mathop{}\!\mathrm{d^#1}}

\hypersetup{
    colorlinks=true,
    linkcolor=red,
    citecolor=blue,
}

\usepackage{amsmath}
\usepackage{amssymb}
\usepackage{subfigure}
\usepackage{hyperref}
\usepackage{url}
\usepackage{xcolor}
\usepackage{color}
\definecolor{amaranth}{rgb}{0.9, 0.17, 0.31}
\definecolor{purple(munsell)}{rgb}{0.62, 0.0, 0.77}
\definecolor{americanrose}{rgb}{1.0, 0.01, 0.24}
\definecolor{palatinateblue}{rgb}{0.15, 0.23, 0.89}
\definecolor{royalblue(web)}{rgb}{0.25, 0.41, 0.88}
\definecolor{hanpurple}{rgb}{0.32, 0.09, 0.98}
\definecolor{beaublue}{rgb}{0.74, 0.83, 0.9}
\definecolor{carminered}{rgb}{1.0, 0.0, 0.22}
\definecolor{brightpink}{rgb}{1.0, 0.0, 0.5}
\definecolor{vividviolet}{rgb}{0.62, 0.0, 1.0}
\hypersetup{ linktoc=all,
    colorlinks, linkcolor={palatinateblue},
    citecolor={brightpink}, urlcolor={amaranth}}

\newcommand{\be}{\begin{equation}}
\newcommand{\ee}{\end{equation}}
\newcommand{\bs}{\begin{split}} 
\newcommand{\bea}{\begin{eqnarray}}
\newcommand{\eea}{\end{eqnarray}}

\newcommand{\bes}{\begin{subequations}}
\newcommand{\ees}{\end{subequations}}

\renewcommand{\d}[1]{\ensuremath{\operatorname{d}\!{#1}}}
\renewcommand{\d}[1]{\ensuremath{\operatorname{d}\!{#1}}}

\newcommand{\bo}{\raise-1mm\hbox{\Large$\Box$}}

\newcommand{\bd}{\boldsymbol}

\begin{document}

%\title{On a radiation correspondence between beta decay and black holes}
%\title{Beta particle trajectory}
%\title{Black hole and dynamical Casimir analog for beta decay}
%\title{On radiation from beta decay, black holes and flying mirrors}
%\title{Black hole and flying mirror for beta decay}
%\title{Black mirror analog for beta decay}
\title{Infrared acceleration radiation}
%\title{Acceleration radiation for beta decay}
%\title{Accelerated boundary for beta decay}
%\title{Quantum power for violent acceleration}
%\title{Beta decay acceleration radiation}
%via a black hole remnant analog moving mirror }
%\title{Black hole remnant analog moving mirror model applied to the bremstrahlung radiation during beta decay to find the power and time-dependent angular distribution}
%\title{Radiation from violent acceleration}%
%\title{Acceleration of the electron during beta decay}
%\title{Relativistic quantum radiation from violent acceleration}
%\title{Relativistic quantum acceleration radiation from beta decay}
%\title{Electron mirrors}
%\title{Soft clouds and moving mirrors}
%\title{Undressing in front of a moving mirror}
%\title{Time and temperature of soft clouds, moving mirrors, and inner bremsstrahlung}
%\title{Soft radiation by violent acceleration}
%\title{Acceleration for infrared radiation}
%\title{Infrared emission from an accelerated electron}
%\title{Infrared emission from a rapidly accelerated relativistic electron}
\author{Michael R.R. Good}
\email{michael.good@nu.edu.kz}
\affiliation{Department of Physics \& Energetic Cosmos Laboratory, Nazarbayev University,
Kabanbay Batyr Ave 53, Nur-Sultan 010000, Qazaqstan}
\author{Paul C.W. Davies}
\email{paul.davies@asu.edu}
\affiliation{Department of Physics and Beyond: Center for Fundamental Concepts in Science,
Arizona State University, Tempe, Arizona 85287, USA}
%\author{Stephen A. Fulling}
%\email{fulling@math.tamu.edu}
%\affiliation{Department of Physics \& Department of Mathematics, Texas A\&M University, College Station, Texas 77843, USA}

\begin{abstract} 
We present an
exactly soluble electron trajectory that permits an analysis of the soft (deep infrared) radiation emitted, the existence of which has been experimentally observed during beta decay via lowest order inner bremsstrahlung. Our treatment also predicts the time evolution and temperature of the emission, and possibly the spectrum, by analogy with the closely related phenomenon of the dynamic Casimir effect.
%We determine the acceleration responsible for soft radiation.  The equation of motion corresponds to deep infrared emission, experimentally observed via lowest order inner bremsstrahlung during beta decay. Corroborating universality, novel predictions reveal time evolution and temperature. 

%n beta decay, the violent acceleration of the electron is unknown.  We use the quantum power of a black hole analog model (dynamical Casimir effect) to determine the trajectory.  
%A novel prediction is the evolution of the distribution and power. 

%In beta decay, the violent acceleration of the electron is unknown.  We use a black hole analog model (moving mirror) to determine the trajectory.  The result agrees with experiment confirming soft photon emission.  A novel prediction is the power evolution. 
%New results are testable, with specific predictions about distribution and power. 
% There are similarities between the radiation emitted by moving mirrors (black hole analogs) and moving point charges.  The scaling on final speed is identical for the total energy emitted during classical beta decay and quantum radiation from an accelerated mirror (associated with a black hole remnant). We investigate the correspondence.   

\end{abstract} 

%\pacs{04.62.+v, 03.67.Hk, 04.70.-s}
\date{\today} 

\maketitle

\begin{figure}[htbp]
\centering
%\begin{subfigure}{0.5\textwidth}
  \centering
  \includegraphics[width=0.8\linewidth]{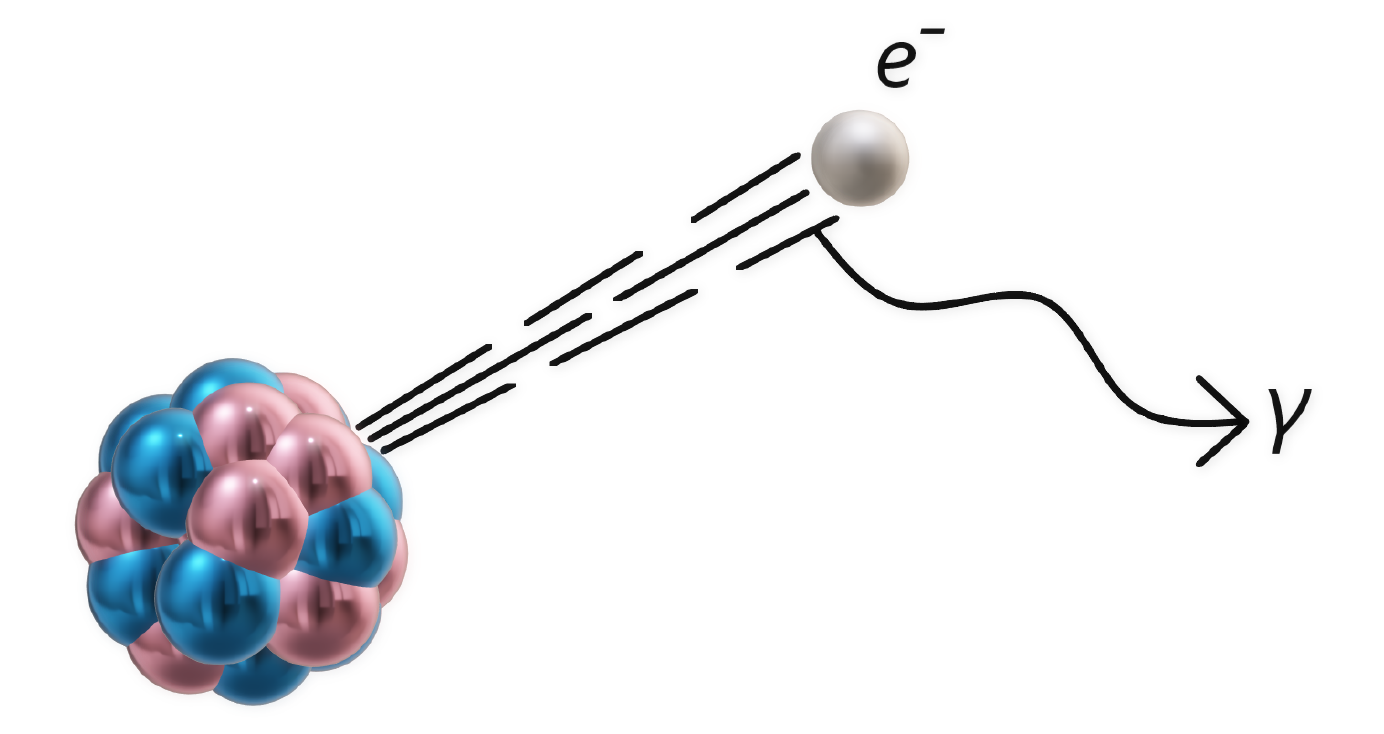}
  %\put(-170,145){a)}
  %\caption{A subfigure}
%\label{fig:4.2a}
%\end{subfigure}%
 %\begin{subfigure}{0.5\textwidth}
%\centering
%\includegraphics[width=0.8\linewidth,angle=90]{UA_spacetime.pdf}
%\put(-180,120){}
  %\caption{A subfigure}
 %\label{fig:sub2}
%\end{subfigure}
\caption{Classical electrodynamics describes the origin story of acceleration radiation as emitted by the electron, known as inner bremsstrahlung (IB). Soft emission is the dominant contribution to the total energy radiated.}
\label{fig1}
\end{figure}

\textit{Introduction. }-The transmutation of a nucleus via beta decay involves the abrupt creation of an electron or positron,
followed by its expulsion from the nucleus. Although this process can be fully understood only in the
context of quantum field theory, there is a long history of classical treatments \cite{bloch}. Viewed classically, beta
decay involves the sudden appearance of a charged particle, which has been modeled by assigning a
step function trajectory to a classical charge \cite{PhysRev.76.365}. The resulting acceleration might be expected to
produce electromagnetic radiation, see Figure \ref{fig1}, and indeed, such radiation has been observed \cite{Ballagh:1983zr}. The process of
photon production accompanying beta decay is sometimes referred to as ‘inner bremmstrahlung (IB).'

The use of a step function is unrealistic, but convenient mathematically \cite{Jackson:490457}. Fortunately, there is a
smoother acceleration function that nevertheless permits an exact treatment of the radiation emission,
and we give that treatment here. By extending the period of acceleration being modeled, we can make a
connection with the well-known Davies-Fulling-Unruh effect \cite{Davies:1974th,Fulling:1972md,unruh76}: in the frame of the charged particle, there
is a thermal bath of photons with a temperature proportional to acceleration. Closely related is the
emission of quanta by an accelerating mirror (moving mirror radiation) \cite{moore1970quantum,DeWitt:1975ys,Davies:1976hi,Davies:1977yv} and the correspondence to black hole radiation \cite{Hawking:1974sw}. The interconnection of charged
particle acceleration and the above mentioned quantum field theory effects has been the subject of
much investigation. In this paper we will not attempt to review these linkages at a fundamental level,
but instead we use the known results phenomenologically to extend the discussion of inner
bremmstrahlung.

%\section{Acceleration Problem}\label{sec2}
%For IB in beta decay, classical electrodynamics bears witness to the origin of the acceleration radiation \cite{PhysRev.76.365}.  The energy of the classical radiation is very small relative to the total energy released but nevertheless has been observed to great accuracy \cite{Ballagh:1983zr}, providing important insights \cite{Ivanov} \cite{Boehm}. 

\textit{Step function example. }-If the electron is initially at rest and imagined to be instantaneously accelerated to a final constant speed, $s= |\vec{\beta}_{\textrm{f}}|$  %|\vec{v}_{\textrm{f}}|/c$ 
where $0<s<1$, then (see e.g. \cite{Zangwill:1507229}),
\begin{equation}
  v(t) =
    \begin{cases}
      s, &  t>0.\\
      0, &  t<0.
     \end{cases}   \label{stepspeed}   
\end{equation}
Working with unit charge, the angular differential distribution of radiated energy is found to be \cite{Jackson:490457}:  
%\be \frac{d^2 I}{d\omega d\Omega} =\frac{e^2}{4\pi^2} \frac{ s^2 \sin ^2\theta }{(1- s \cos\theta)^2}.\label{angelectron1}\ee
\be \frac{d^2 E}{d\omega d\Omega} =\frac{1}{16\pi^3} \left(\frac{ s \sin \theta }{1- s \cos\theta}\right)^2,\label{angelectron1}\ee
where $\theta$ is the angle between the final velocity $\vec{\beta}_{\textrm{f}}$ and the observation point of the radiation. Integration of Eq.~(\ref{angelectron1}) over solid angle $d\Omega = \sin\theta \d\theta \d \phi$ and over frequencies IR/UV-limited by cutoffs $\Delta_\omega\equiv \omega_{\textrm{max}}-\omega_{\textrm{min}}$ gives the energy radiated by the electron. The total energy is rendered finite in this interval, 
%\be E_{\textrm{electron}} = \frac{\omega_{\textrm{c}}}{2\pi^2} \left( \frac{\eta}{s} - 1\right).\ee
\be E =  \frac{1}{4\pi^2} \left[\frac{1}{s}\ln \left(\frac{1+s}{1-s}\right)-2\right]\Delta_\omega.\label{EnergyElectron1} \ee
%\bea \nonumber E &=& \frac{q^2}{\pi} \left[\frac{1}{s}\ln \left(\frac{1+s}{1-s}\right)-2\right]\omega_{\textrm{c}},\\
%E &=& \frac{q^2}{\pi} \left( \frac{\eta}{s} - 1\right)2\omega_{\textrm{c}}.\label{EnergyElectron1}\eea
%\be E_{\textrm{electron}} = \frac{2e^2}{\pi \epsilon_0 c} \left( \frac{\eta}{s} - 1\right)\omega_{\textrm{c}}.\label{EnergyElectron}\ee
The detector sets the energy scale sensitivity. Eq.~(\ref{EnergyElectron1}) is lowest order IB energy \cite{PhysRev.76.365}, and has been observed to great accuracy \cite{Ballagh:1983zr}.  The foregoing treatment is sometimes referred to as the instantaneous collision formalism \cite{Cardoso:2002pa,Cardoso:2003cn}.

%This 1949 computation by Chang and Falkoff \cite{PhysRev.76.365} is treated in Jackson \cite{Jackson:490457} and Zangwell \cite{Zangwill:1507229}. 
%Also called the instantaneous collision formalism, Eq.~(\ref{EnergyElectron1}) has been used to help compute the gravitational energy released during the quantum creation of pairs of black holes \cite{Cardoso:2002pa} and the electromagnetic radiation released during the high energy collision of a rapidly accelerated charged point particle with a black hole \cite{Cardoso:2003cn}. 

Not only is it physically desirable to avoid the infinite acceleration of Eq.~(\ref{stepspeed}), but the mathematical use of the discrete step velocity limits the final results to quantities independent of time. The radiated energy Eq.~(\ref{EnergyElectron1}), is characterized by universality and a classical limit from a corresponding time-dependent trajectory \cite{bloch}. Knowing the continuous acceleration responsible for deep IR could help provide a simple underlying physical connection to gravitation via the Equivalence Principle.%We remedy these deficiencies by providing an analytic and continuous function.  

\begin{figure}[htbp]
\centering
%\begin{subfigure}{0.5\textwidth}
  \centering
  \includegraphics[width=1.0\linewidth]{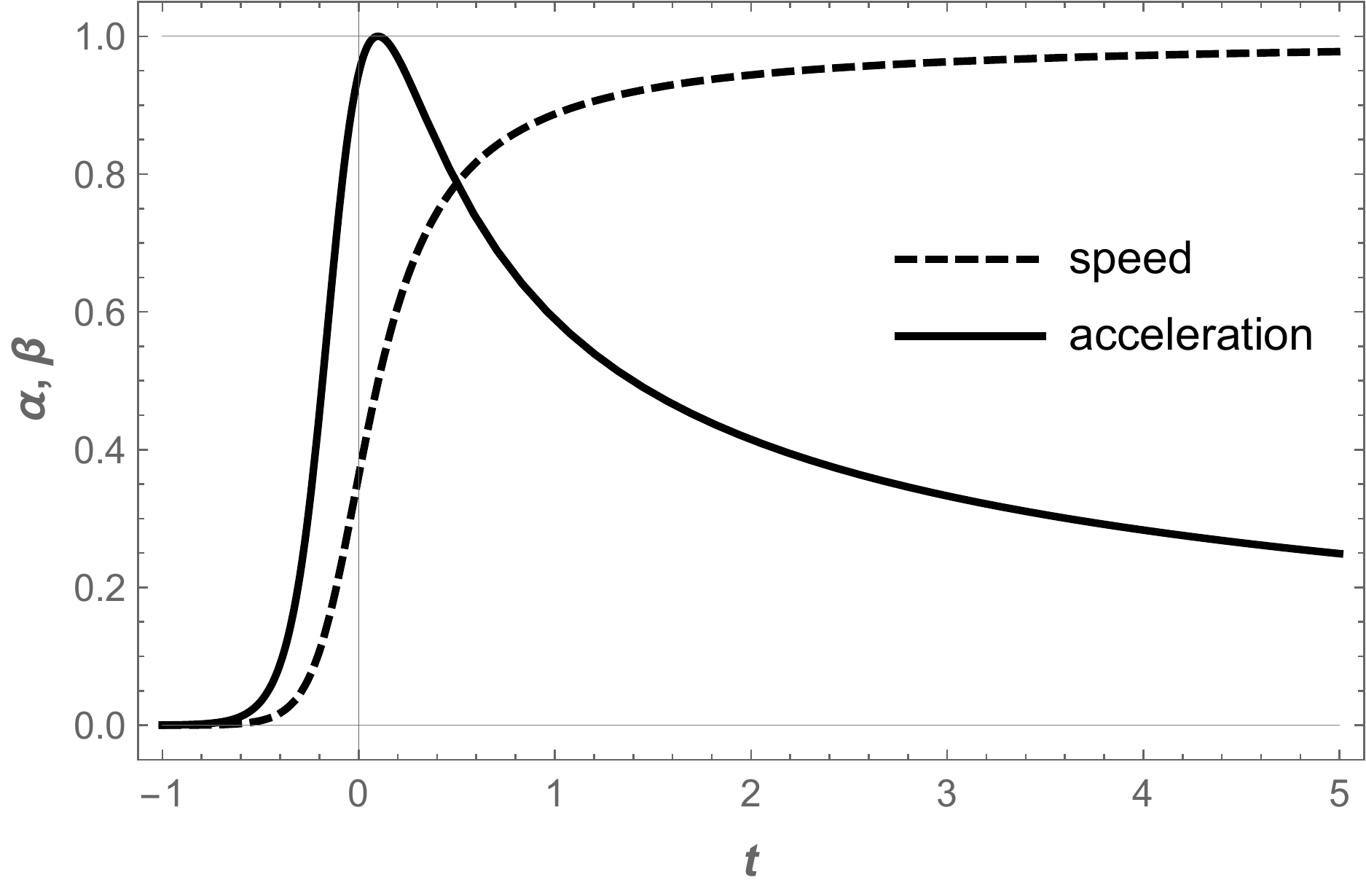}
 \caption{The proper acceleration is always finite, continuous, and has a maximum after but near $t=0$ (for illustration the acceleration has been normalized by its maximum).  For large $\kappa$, the speed approaches the step-function form of Eq.~(\ref{stepspeed}) and the smooth acceleration approaches a delta-function.  Here $\kappa = 10$, $s=0.999$.}
\label{fig2}
\end{figure}

\begin{figure}[htbp]
\centering
%\begin{subfigure}{0.5\textwidth}
  \centering
  \includegraphics[width=0.9\linewidth]{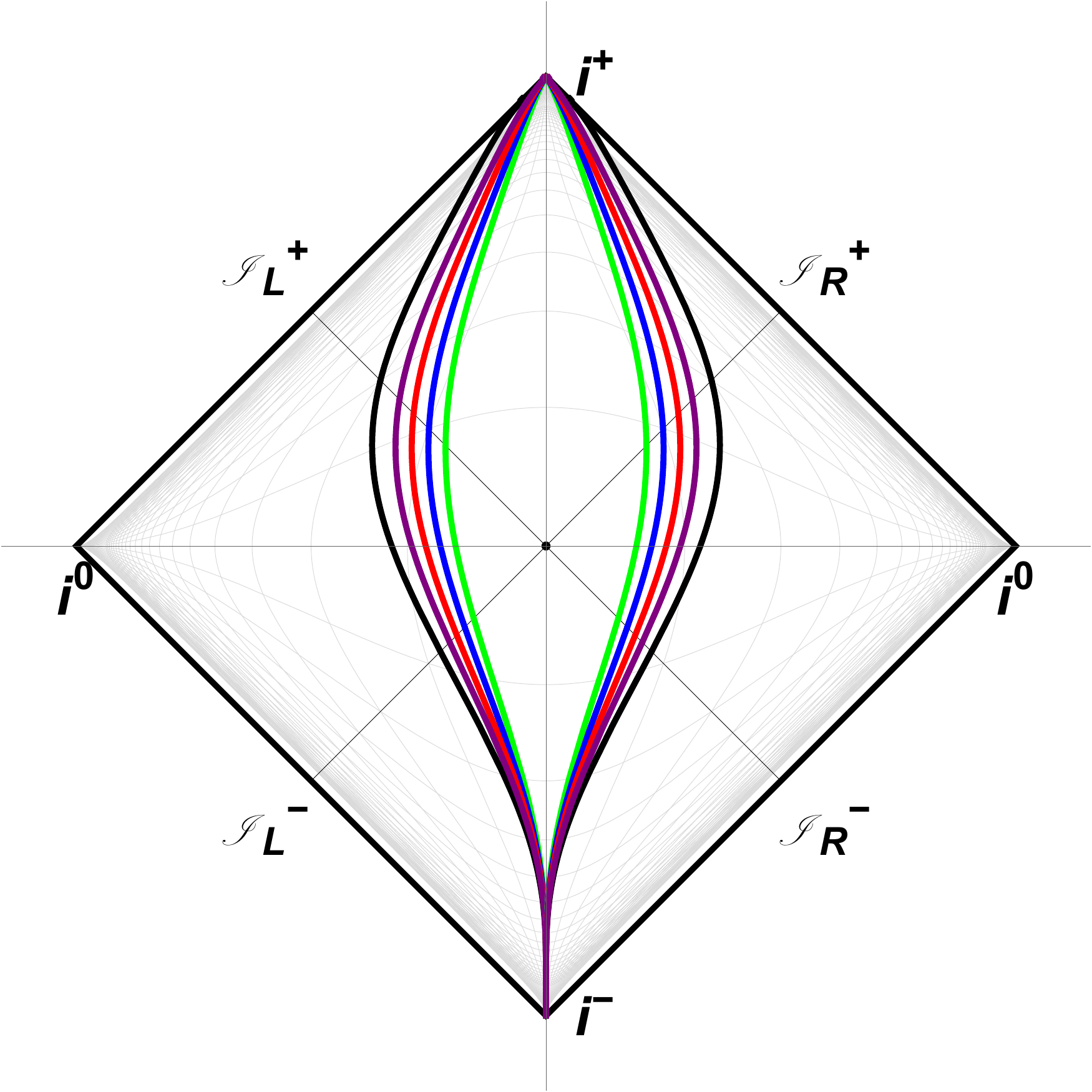}
 \caption{Penrose diagram of the trajectory class, Eq.~(\ref{invspeed1}) demonstrating time-like asymptotic inertia, $\kappa =1$.  For left-right visual clarity, we have plotted in (1+1) dimensions of spacetime. The trajectory starts with asymptotic zero velocity and finishes with asymptotic constant velocity. The power, Eq.~(\ref{powerspeed1}), is independent of whether the electron moves to the right or left (depicted), but the angular distribution, Eq.~(\ref{angelectron1}), picks up a sign on the final speed when moving to the left. Here from inside-out, the final speeds are $s=0.55,0.65,0.75,0.85,1.00$. }
\label{fig3}
\end{figure}

\textit{Smooth acceleration. }-Under the above motivations, we consider the trajectory,
\be \frac{dt}{dr} = \frac{1}{\kappa  r}+\frac{1}{s}. \label{invspeed1}\ee
The asymptotic speeds are $v = (0,s)$ as $r\to (0,\infty)$ (the electron moves to the right by convention\footnote{In the closely related moving mirror model, (see \cite{Good:2016yht}), the usual convention is to move to the left (see Figure \ref{fig3}). The difference is a sign change in the angular distribution. The energy remains invariant.} \cite{Jackson:490457}), matching Eq.~(\ref{stepspeed}). The proper acceleration, $\alpha = d\gamma/dr$, has time-dependence, $\alpha(t) = \kappa\beta\gamma^3(1-\beta/s)^2$, and possesses asymptotic inertia. See Figure \ref{fig2} \& \ref{fig3} for illustration.  Here $\kappa$ is the dimensionful acceleration parameter of the model which corresponds to the sensitivity in frequency range of the detector and sets the scale: $\kappa 
\leftrightarrow 12 \Delta_\omega/\pi$. With large $\kappa$, the speed of Eq.~(\ref{invspeed1}) approaches the step function example, Eq.~(\ref{stepspeed}). However, no such approximation is needed to obtain Eqs.~(\ref{angelectron1}) or (\ref{EnergyElectron1}).    

\textit{Time-distribution \& power. }-The time-dependent power distribution is computed using Eq.~(\ref{invspeed1}) with straightforward vector algebra (see the procedure in \cite{Good:2019aqd}), 
\be \frac{dP}{d\Omega} = \frac{\kappa ^2 \beta ^2 (\beta -s)^4\sin ^2\theta }{16 \pi^2  s^4 (1-\beta  \cos \theta)^5},\label{shape1}\ee
where $s$ again, is the final constant speed, and $\beta = \beta(t)$ is the time-dependent velocity. Integration over time, gives the time-independent angular differential distribution of energy, which turns out to be identical to result for the step-function trajectory, Eq.~(\ref{angelectron1}). %Integration over time, gives the time-independent angular differential distribution of energy, Eq.~(\ref{angelectron1}). This confirms that the proposed trajectory, Eq.~(\ref{invspeed1}), yields the identical distribution in space of radiated energy, supporting the notion that Eq.~(\ref{invspeed1}) is a good candidate for the unbroken motion.  

Moreover, using the Lorentz-invariant proper acceleration in $P=\alpha^2/6\pi$, we obtain the total power radiated,
\be P = \frac{\kappa^2 \gamma^6\beta ^2}{6\pi} \left(1-\frac{\beta}{s}\right)^4 .\label{powerspeed1}\ee
The total radiated energy for the entire trajectory is readily obtained by integrating Eq.~(\ref{powerspeed1}) over time,
which again yields an identical result to the step-function case, given by Eq.~(\ref{EnergyElectron1}). The fact that the more
realistic smooth trajectory recapitulates the earlier results justifies the use of our choice of Eq.~(\ref{invspeed1}). However, our model has the advantage that we can examine the behavior of the accelerated charge over time.

%Integration of Eq.~(\ref{powerspeed1}), over space with the appropriate Jacobian, $dt/dr$ of Eq.~(\ref{invspeed1}), and correct bounds, $(r; 0,\infty)$, gives the total energy, Eq.~(\ref{EnergyElectron1}). This is additional confirmation that the acceleration is appropriately modeled by trajectory Eq.~(\ref{invspeed1}). 

\begin{figure}[htbp]
\centering
%\begin{subfigure}{0.5\textwidth}
  \centering
  \includegraphics[width=0.9\linewidth]{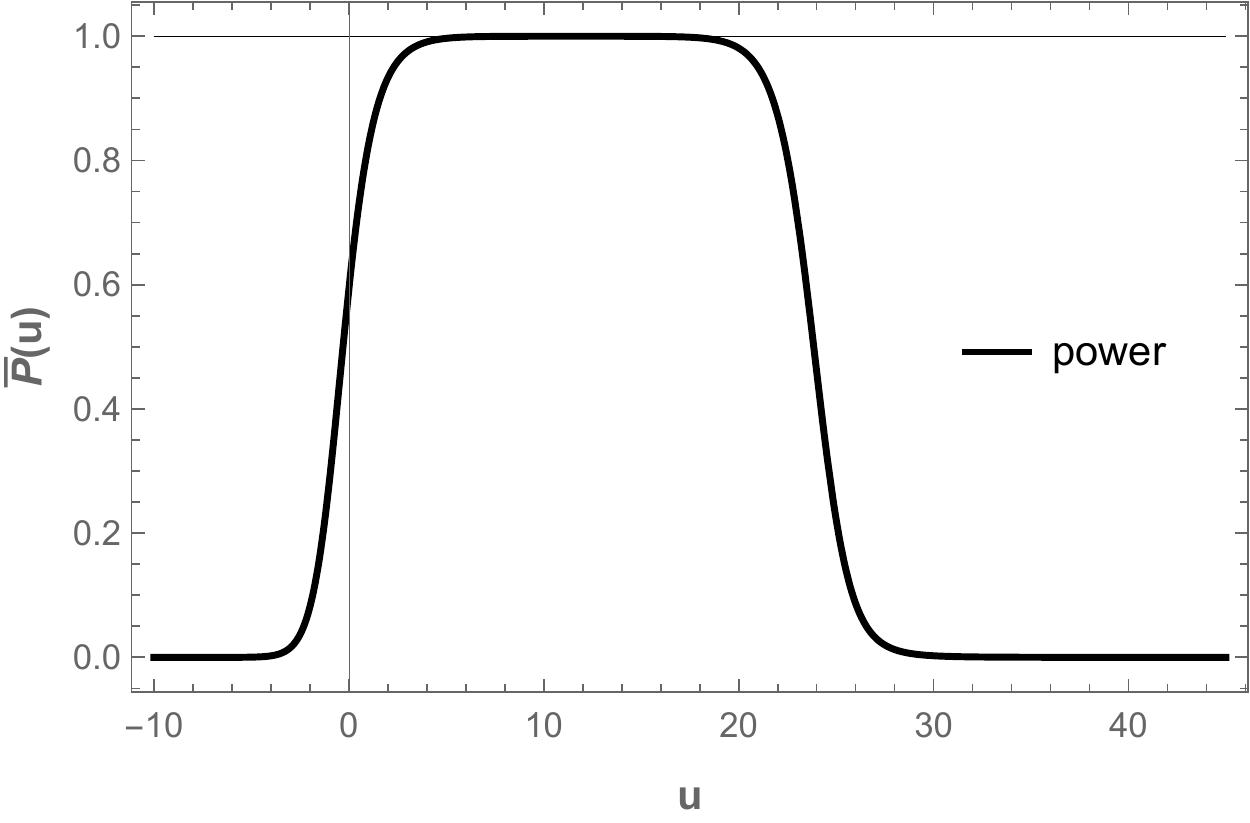}
 \caption{A plot of the power, $\bar{P}(u)$, with a plateau demonstrating constant emission when the final speed of the electron is extremely ultra-relativistic, $s=1-10^{-11} = 0.99999999999$. Here $\kappa =1$. This plateau corroborates the conclusion that at high electron speeds the photons find themselves in a Planck distribution, Eq.~(\ref{betaT1}) with temperature $T=\kappa/2\pi$, Eq.~(\ref{temp}). The vertical scale has been normalized by $\kappa^2/48\pi$ so that the plateau is at height $\bar{P}(u) =1$.  The integral under the curve, Eq.~(\ref{barP}), is the experimentally observed soft IB energy, Eq.~(\ref{EnergyElectron1}).  }
\label{fig5}
\end{figure}

\textit{Equilibrium emission. }-Interestingly, a period of constant emission is present in the power measured by a far away observer.  Best represented as %We can corroborate the temperature with another useful measure of power as 
the change of energy with respect to retarded time $u=t-r$, and written as $\bar{P} = \frac{\diff E}{\diff u},$ such that
\be E = \int_{-\infty}^{\infty} \bar{P}(u) \diff u, \label{barP}\ee
we write $\bar{P} = P \frac{\diff t}{\diff u} = P/(1-v)$.  Formulating $\bar{P}(u)$ in terms of retarded time, gives a lengthy result, but we plot the measure $\bar{P}(u)$ at high final asymptotic speeds $s \sim 1$ and reveal a constant power plateau indicative of thermal emission. Additionally, beta Bogolubov coefficients corroborates this radiative equilibrium via an explicit Planck distribution in Eq.~(\ref{betaT1}). See a plot of the power plateau in Figure \ref{fig5}.

\textit{Radiation reaction. }-Having computed the  power, $P=\alpha^2/6\pi$, we now turn to the self-force, $F=\alpha'(\tau)/6\pi$.  It is analytically tractable, and a concise expression is given in terms of speed $\beta$, 
\be F = \frac{\kappa^2 \gamma^6 \beta^2}{6\pi}\left(1-\frac{\beta }{s}\right)^3 \left(2 \beta +\frac{1}{\beta }-\frac{3}{s}\right). \label{softself}\ee
The self-force is zero at maximum power.  Integrating over distance gives the work done, 
\be W = \int_0^\infty F(r) \diff r = -\int_{-\infty}^\infty P(t) \diff t = -E. \ee That is, taking Eq.~(\ref{softself}) over $d\beta$ using, $d\beta/dr =  \kappa (1-\beta/s)^2$, where $\beta$ ranges from $(0,s)$, one obtains the energy associated with the self-force.  The resulting work is $W=-E$, the equal and opposite of Eq.~(\ref{EnergyElectron1}). This demonstrates consistency between the radiation reaction and conservation of energy. 

\textit{Universality, spectra \& temperature. }-The preceding results derived in the context of IB are the same for the scattering of Faddeev-Kulish electrons in QED where a cloud of soft photons exist in the dressed state \cite{Tomaras}. Moreover, the same results hold true for the perfectly reflecting moving mirror \cite{Good:2016yht} of the dynamical Casimir effect.  In turn, the accelerated boundary correspondence between mirrors and black holes \cite{ Good:2019tnf}, demonstrates trajectory Eq.~(\ref{invspeed1}) induces an exact analog of black hole evaporation leading to a remnant \cite{wilczek1993quantum}. The unexpected synthesis of IB, clouds, mirrors, and remnants corroborate the universality of the deep infrared.  

\begin{figure}[htbp]
\centering
%\begin{subfigure}{0.5\textwidth}
  \centering
  \includegraphics[width=1.0\linewidth]{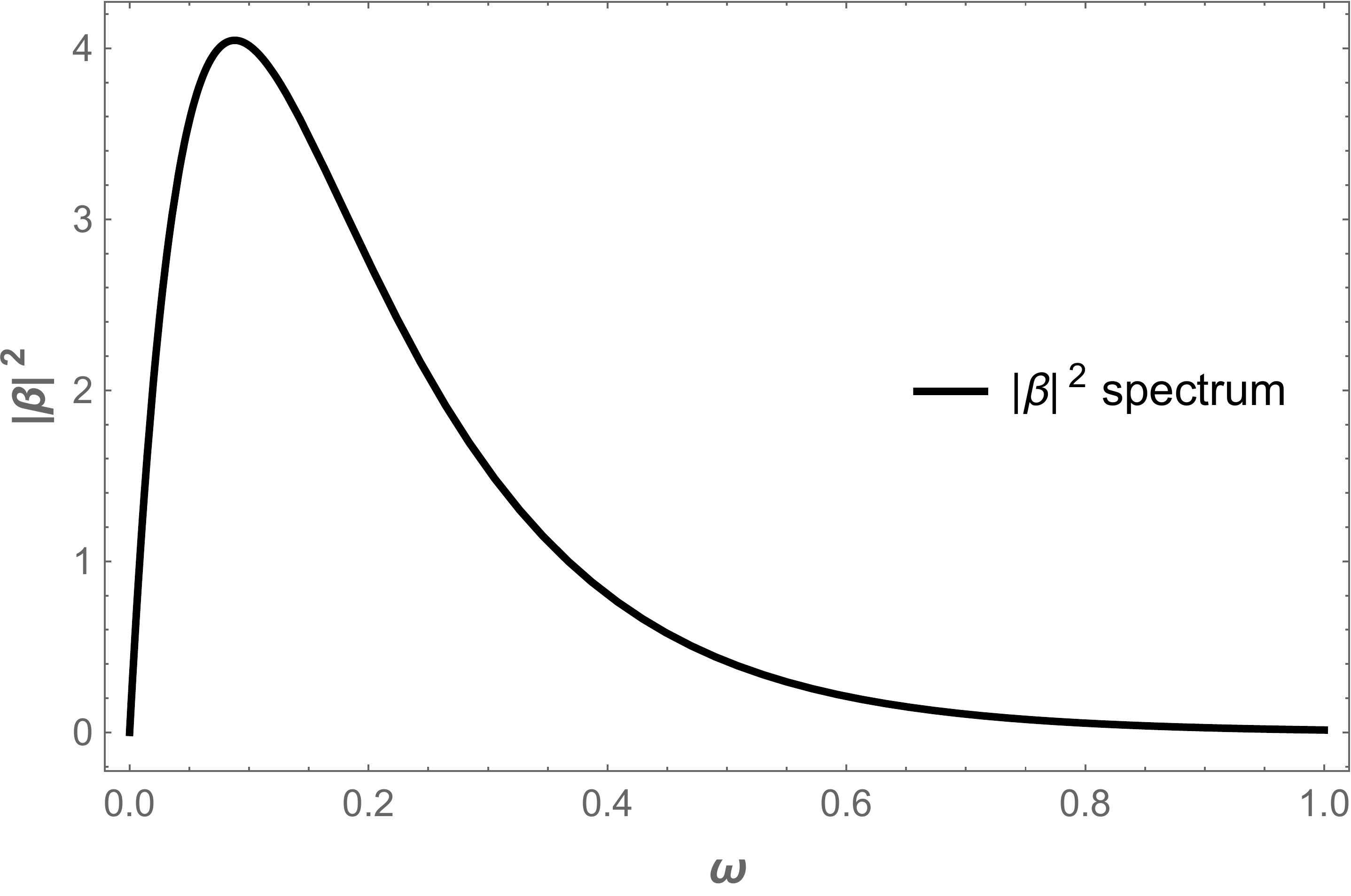}
 \caption{The $|\beta_{\omega\omega'}|^2$ spectrum of Eq.~(\ref{betaT1}). Here $\omega'=\kappa=1$ and $s=1/2$. The vertical axis has been scaled by $10^5$ for visual clarity. The qualitative black-body shape is indicative of the explicit Planck factor in Eq.~(\ref{betaT1}).}
\label{fig4}
\end{figure}

Since accelerating boundaries radiate soft particles whose long wavelengths lack the capability to probe the internal structure of the source \cite{Strominger:2017zoo}, we compute, in the spirit of analogy, the moving mirror spectrum (scaled by $\kappa$) as an illustration of what the soft-spectrum for IB might look like.  Combining the results for each side of the mirror \cite{Good:2016yht} by adding the squares of the beta Bogolubov coefficients, the overall spectrum is
\be |\beta_{\omega\omega'}|^2 = \frac{2 \omega  \omega ' \left(\omega _{\bar{s}}^2+\omega _s^2\right)}{\pi  \kappa  \omega _s^2 \omega _T \omega _{\bar{s}}^2 \left(e^{\frac{2 \pi}{\kappa}  \omega _T}-1\right)}.\label{betaT1}\ee
Here $\omega_s = \left(\frac{1}{s}-1\right) \omega '+\left(\frac{1}{s}+1\right) \omega$, and $\omega_{\bar{s}} = \left(-\frac{1}{s}-1\right) \omega '+\left(1-\frac{1}{s}\right) \omega$. The total frequency is $\omega_T = \omega + \omega'$. A numerical integration of 
\be E = \int_0^\infty \int_0^\infty \omega |\beta_{\omega\omega'}|^2 \d \omega \d \omega', \label{numerical}\ee
confirms the total energy radiated, Eq.~(\ref{EnergyElectron1}).  %Thus, the new expression, Eq.~(\ref{betaT1}), plotted in Figure \ref{fig4}, for the particle spectrum of the radiation, may help characterize the color of the light emitted during beta decay. 
Given the close association between accelerating mirrors and charges, we postulate that the IB spectrum in beta decay is likely to be of the same form as Eq.~(\ref{betaT1}).  We have plotted the spectrum of the moving mirror radiation in Figure \ref{fig4}.  If experiment confirms our prediction, then one could regard soft IB from beta decay as an analogue of the dynamical Casimir effect.  

The explicit Planck factor demonstrates the particles, $N(\omega) = \int d\omega'|\beta_{\omega\omega'}|^2$, are distributed with a temperature, 
\be T=
\frac{\kappa}{2\pi} = \frac{6}{\pi^2} \Delta_\omega,\label{temp}\ee in the high frequency approximation $\omega'\gg\omega$ \cite{Hawking:1974sw}. Recall $\Delta_\omega\equiv \omega_{\textrm{max}}-\omega_{\textrm{min}}$, is the scale set by the sensitivity of detection.  Thermal emission is not surprising considering the power plateau (Figure \ref{fig5}) and the close analogy for quantum and classical quantities of powers  \cite{Good:2021ffo,Zhakenuly:2021pfm} and self-forces \cite{Myrzakul:2021bgj,Ford:1982ct} between mirrors and electrons. 
 %In light of these results, soft IB from beta decay is an observational analog of the dynamical Casimir effect.

%Alternatively, the situation of violent acceleration can be viewed, pictured in Figure \ref{fig1}, as the sudden switching on of the charge \cite{Jackson:490457}.  The charge creation happens in the same short time interval. This equivalent viewpoint is quantum mechanical in origin, which is in harmony with the quantum nature of the conformal anomaly responsible for the radiation from the moving mirror.

\textit{Conclusion. }-
We have calculated the deep infrared radiation emitted by a rapidly accelerating classical point charge using a smooth trajectory that permits exact solution of all relevant quantities.  We have derived novel time-dependent power and angular distribution formula. The soft self-force was computed, universality was highlighted across several distinct systems, and Bogolubov coefficient spectra were obtained,  demonstrating consistency with the observed energy.  The temperature of the light is found via a Planck distribution.  The key result, from which the others flow, is an analytic continuous equation of motion for infrared acceleration radiation.

\textit{Acknowledgements. }-We thank Stephen Fulling for useful discussion. Funding comes in part from the FY2021-SGP-1-STMM Faculty Development Competitive Research Grant No. 021220FD3951 at Nazarbayev University. Appreciation is given to the organizers, speakers, and participants of the QFTCS Workshop: May 23-27, 2022, at which preliminary results were first presented and helpful feedback are included therein.       
%\newpage

%\twocolumngrid
%%%%%%%%%%%%%%%%%%%%%%%%%%%%%%%%%%%%% References in order
\bibliography{main} 
\end{document}